\documentclass[showpacs,twocolumn,superscriptaddress]{revtex4}

\usepackage{doi}
\usepackage{hyperref}
\hypersetup{
  colorlinks=true,        
  linkcolor=blue,         
  citecolor=cyan,         
}
\usepackage[utf8]{inputenc}
\usepackage{graphicx}
\usepackage{dcolumn}
\usepackage{bm}
\usepackage{color}
\usepackage{enumitem}
\usepackage{amsmath}
\usepackage{amssymb}
\usepackage{orcidlink}
\usepackage{soul}
\usepackage{color}

\begin{document}

\title{Plasma impact on black hole shadow and gravitational weak lensing for Schwarzschild-like black hole}

\author{Weiqiang Yang,\orcidlink{0009-0002-9333-9128}}
\email{25s011087@stu.hit.edu.cn}
\affiliation{School of Physics, Harbin Institute of Technology, Harbin 150001, People’s Republic of China}

\author{Mirzabek Alloqulov,\orcidlink{0000-0001-5337-7117}}
    \email{malloqulov@gmail.com}
    \affiliation{School of Physics, Harbin Institute of Technology, Harbin 150001, People’s Republic of China}
    \affiliation{University of Tashkent for Applied Sciences, Str. Gavhar 1, Tashkent 100149, Uzbekistan}
    \affiliation{New Uzbekistan University, Movarounnahr str. 1, Tashkent 100000, Uzbekistan}
    
\author{Ahmadjon Abdujabbarov,\orcidlink{0000-0002-6686-3787}}
\email{ahmadjon@astrin.uz}
\affiliation{School of Physics, Harbin Institute of Technology, Harbin 150001, People’s Republic of China}
\affiliation{Institute of Theoretical Physics, National University of Uzbekistan, Tashkent 100174, Uzbekistan}

\author{Bobomurat Ahmedov,\orcidlink{0000-0002-1232-610X}}
\email{ahmedov@astrin.uz}

\affiliation{School of Physics, Harbin Institute of Technology, Harbin 150001, People’s Republic of China}
\affiliation{Institute of Theoretical Physics, National University of Uzbekistan, Tashkent 100174, Uzbekistan}
\affiliation{Institute for Advanced Studies, New Uzbekistan University, Movarounnahr str. 1, Tashkent 100000, Uzbekistan}

\author{Chengxun Yuan,\orcidlink{0000-0002-2308-6703}}
\email{yuancx@hit.edu.cn}

\affiliation{School of Physics, Harbin Institute of Technology, Harbin 150001, People’s Republic of China}

\author{Chen Zhou,\orcidlink{0000-0002-1361-3717}}
\email{chenzhou@hit.edu.cn}

\affiliation{School of Physics, Harbin Institute of Technology, Harbin 150001, People’s Republic of China}

%
\date{\today}
\begin{abstract}
{
This article delves into the observational properties of a Schwarzschild-like black hole (BH). Initially, the research provides a succinct examination of the spacetime geometry and the configuration of its horizon. Furthermore, we study the photon dynamics around the Schwarzschild-like BH in the presence of the plasma using the Hamiltonian formalism. It was found that the photon sphere radii increase under the influence of the plasma frequency and vice versa for the spacetime parameters. Further exploration is dedicated to understanding how the plasma affects the shadow of the BH, and we find that the radius of the BH shadow shrinks with the rise of the $\xi$ parameter and plasma frequency. We then turn to the getting constraint of the spacetime parameters and the plasma frequency by using the observational data released by the Event Horizon Telescope (EHT) collaboration for the M87* and Sgr A*. Additionally, the research scrutinises the phenomenon of gravitational weak lensing in the vicinity of a Schwarzschild-like BH, considering both uniform and non-uniform plasma scenarios. The outcomes demonstrate that the angle of deflection increases under the influence of a uniform plasma frequency, whereas the opposite is true for non-uniform plasma. In both scenarios, a rise in the spacetime parameters results in a decrease in the deflection angle. Finally, we investigate the magnification of the gravitationally lensed image. The effect of the spacetime parameters and plasma frequencies on the total magnification are same as in the deflection angles. }
\end{abstract}

\maketitle

\section{Introduction}

The past decade has witnessed remarkable progress in black hole (BH) astrophysics, transforming these objects from purely theoretical constructs into observable laboratories for testing the gravity in strog-field regime. The groundbreaking first imaging of the supermassive BHs M87* and Sgr A* by the Event Horizon Telescope (EHT) collaboration~\cite{Akiyama2019a,Akiyama2019b,Akiyama2022} provided direct visual evidence of BH shadows, while the detection of gravitational waves from the binary BH mergers by LIGO and VIRGO~\cite{Abbott2016} opened a new window for probing the dynamics of compact objects. Aforementioned observational breakthroughs stimulated extensive theoretical investigations aimed at constraining deviations from the Kerr paradigm and testing alternative theories of gravity~\cite{DuranCabaces2025,Guerrero2021,Olmo2023,Gralla2019}. 

Despite the empirical successes of classical general relativity (GR), fundamental theoretical issues remain unresolved. The singularity theorems established by Roger Penrose~\cite{Penrose1965} demonstrate that, under reasonable physical assumptions, the gravitational collapse inevitably leads to the formation of singularities where curvature invariants diverge, and classical physics breaks down. This indicates that GR is incomplete at extreme energy scales, necessitating the incorporation of quantum gravity effects to resolve these pathological features.

Among the various approaches to quantum gravity, the asymptotic safety scenario~\cite{Reuter1998,Niedermaier2006, Reuter2019} emerged as a compelling framework. This approach proposes that gravity is non-perturbatively renormalizable due to the existence of a non-trivial ultraviolet fixed point that governs the high-energy behaviour of the theory. The functional renormalization group provides a systematic tool to investigate the running of gravitational couplings with the energy scale, revealing a characteristic scale dependence of Newton's constant. A powerful phenomenological method to incorporate quantum gravity effects into classical spacetime geometries is the renormalization group improvement procedure~\cite{Bonanno2000,Bonanno2006,Falls2018}. This technique utilizes the running of couplings to construct effective metrics that interpolate between the classical infrared regime and the quantum-dominated ultraviolet regime. Recently, the authors of Ref.~\cite{Alencar2026} constructed an explicit renormalization group improved Schwarzschild-like BH spacetime. This new solution exhibits a scale-dependent Newton constant $G(r)$ that interpolates between the classical value $G_0$ at large distances and a running coupling in the ultraviolet, yielding a regular spacetime with a de Sitter-like core while preserving the asymptotic Schwarzschild structure.

An essential aspect of testing such geometries lies in understanding their observational signatures, which are the BH shadow and weak gravitational lensing. In realistic astrophysical environments, BHs are typically surrounded by various forms of matter, among which plasma constitutes a ubiquitous component. The presence of plasma significantly modifies the propagation of electromagnetic waves, introducing a frequency-dependent effective mass for photons and altering their trajectories compared to vacuum propagation~\cite{Synge1960, Perlick2004, Tsupko2012}. This plasma-induced modification of the geodesic structure provides an additional probe of both the background spacetime and the properties of the surrounding medium~\cite{BisnovatyiKogan2010}. To this day, there are different types of studies that have investigated the BH shadow and gravitational weak lensing around the compact objects~\cite{Perlick15a,Synge:1960b,Bozza2002,Rog:2015a,Morozova2013,Abdujabbarov2015,Alloqulov20251,Al-Badawi2024,Alloqulov_2023,Jiang2024,Alloqulov_2024,Al-Badawi20242,Alloqulov20242,Atamurotov2022,Sharipov2026461,Alloqulov20252,Khasanov_2025,Javed:2022fsn,Atamurotov:2022knb}. In this work, we aim to study the photon dynamics around the Schwarzschild-like BH~\cite{Alencar2026} surrounded by plasma using the Hamiltonian formalism, together with the BH shadow and the gravitational weak lensing.

The structure of this paper is organized in the following way. In Section~\ref{2}, we review the spacetime of the Schwarzschild-like BH, including the event horizon structure. In addition, we investigate the motion of a photon around the BH surrounded by plasma using the Hamiltonian formalism. The impact of the plasma on the radii of the photon sphere and the BH shadow was explored in this part, together with the constraint values of the spacetime parameters and plasma frequency by using the EHT collaboration results. The weak gravitational lensing and the magnification of the gravitationally lensed images were studied in Sections~\ref{3} and~\ref{4}, respectively. Finally, we summarize our conclusions and discussions in Section~\ref{con}. Throughout this work, we adopt natural units with $c=\hbar=1$ and the metric signature $(-,+,+,+)$.

\section{Plasma impact on black hole shadow}\label{2}
\begin{figure*}
    \centering
    \includegraphics[scale=0.42]{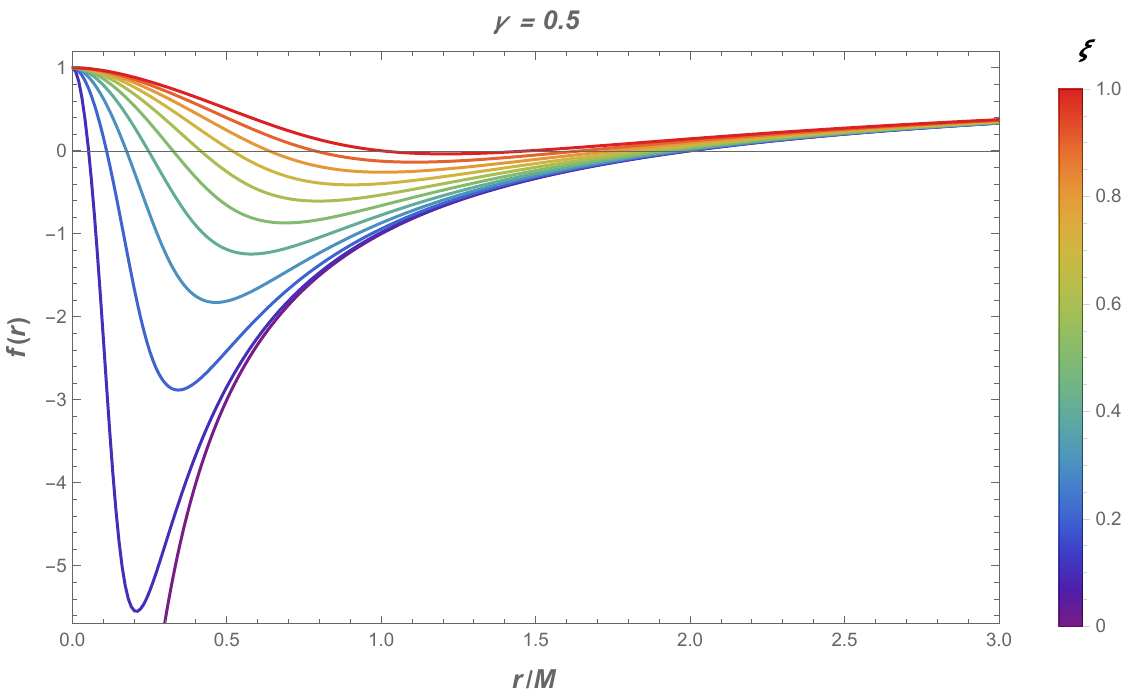}
     \includegraphics[scale=0.42]{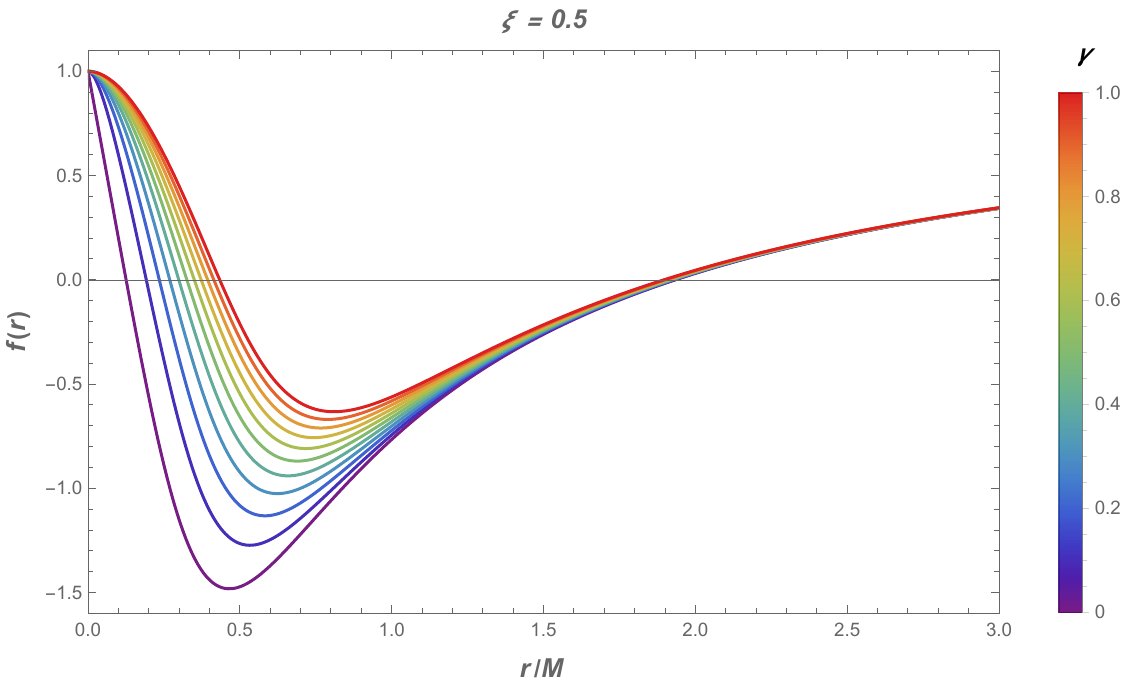}
    \caption{{The plot shows the radial dependence of the metric function for the different values of the $\xi$ parameter (left panel) and $\gamma$ parameter (right panel). Here, we set $\gamma=0.5$ and $\xi=0.5$ for the left and right panels, respectively.}}
    \label{fig:fr}
\end{figure*}

\subsection{Photon dynamics around black hole surrounded by plasma}

\begin{figure*}
    \centering
    \includegraphics[scale=0.55]{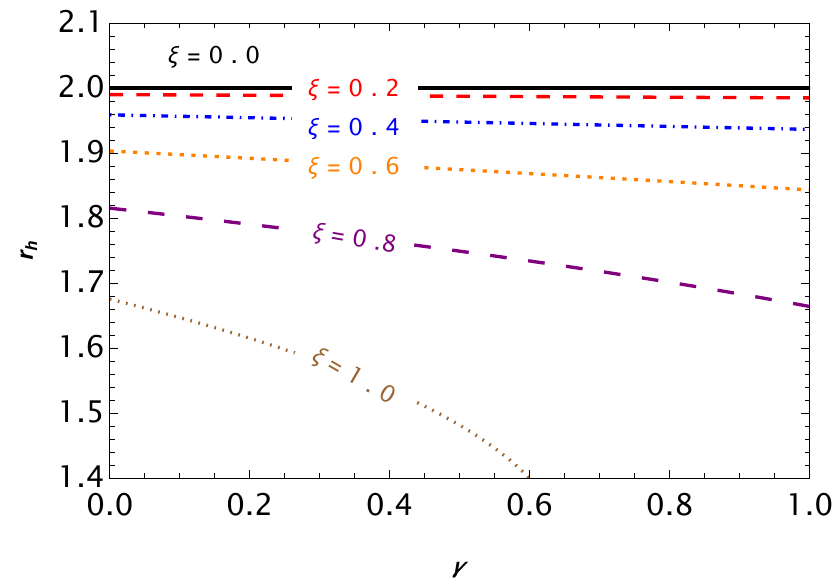}
    \includegraphics[scale=0.55]{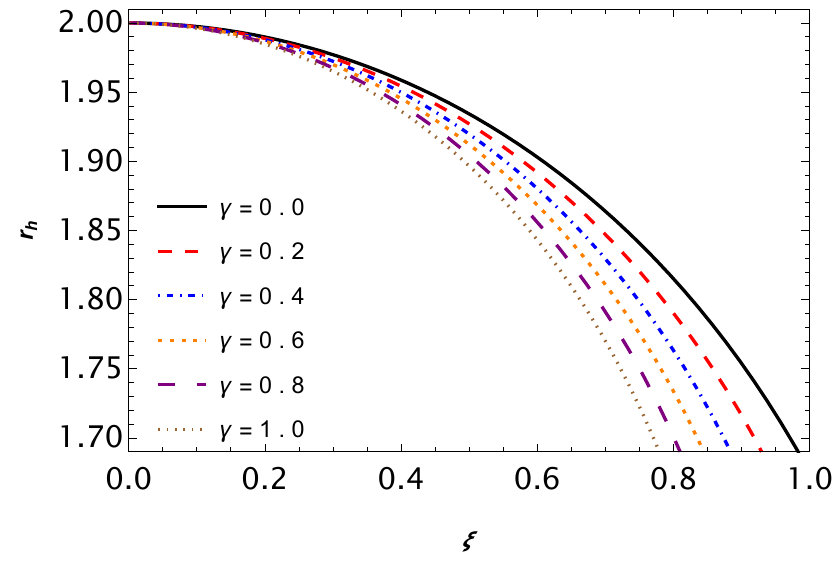}
    \caption{{The plot illustrates the dependence of the event horizon radii of the Schwarzschild-like BH on the spacetime parameters.}}
    \label{rphoton}
\end{figure*}

{We consider the spacetime of the Schwarzschild-like BH in the following form
\begin{equation}
 ds^2=-f(r)dt^2+\frac{dr^2}{f(r)}+r^2(d\theta^2+\sin^2{\theta}d\phi^2)\,,   
\end{equation}
where 
\begin{equation}
    f(r)=1-\frac{4Mr^2}{\xi^2(\gamma M+r)+\sqrt{\xi^4(\gamma M+r)^2+4r^6}}\,.
\end{equation}
The metric function includes the $\xi$ and $\gamma$ parameters, which refer to the cutoff scale and interpolation parameters, respectively. It is worth noting that we can recover the standard Schwarzschild spacetime when $\xi$ parameter tends to zero (see, Ref.~\cite{Bonanno2000}). In Fig.~\ref{fig:fr}, we plot the radial dependence of the metric function for the different values of the spacetime parameters. By solving the $f(r)=0$, we can find the event horizon radii, and we plot the radius of the event horizon as a function of the spacetime parameters in Fig.~\ref{rphoton}. It can be seen from this figure that the event horizon radii decrease under the influence of both $\xi$ and $\gamma$ parameters.
}

\begin{figure*}[!htb]
 \begin{center}
   \includegraphics[scale=0.5]{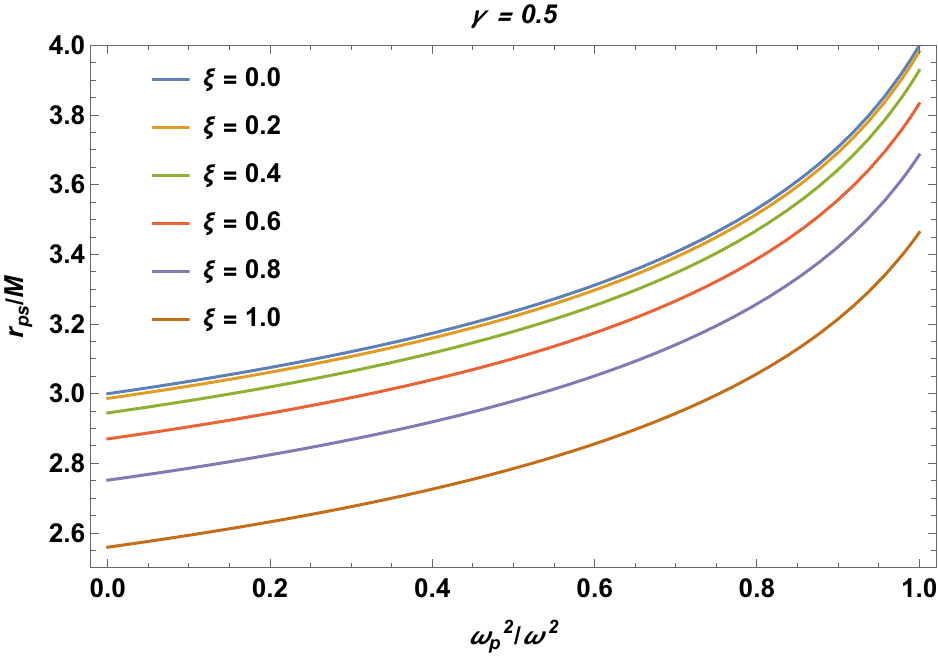}
   \includegraphics[scale=0.5]{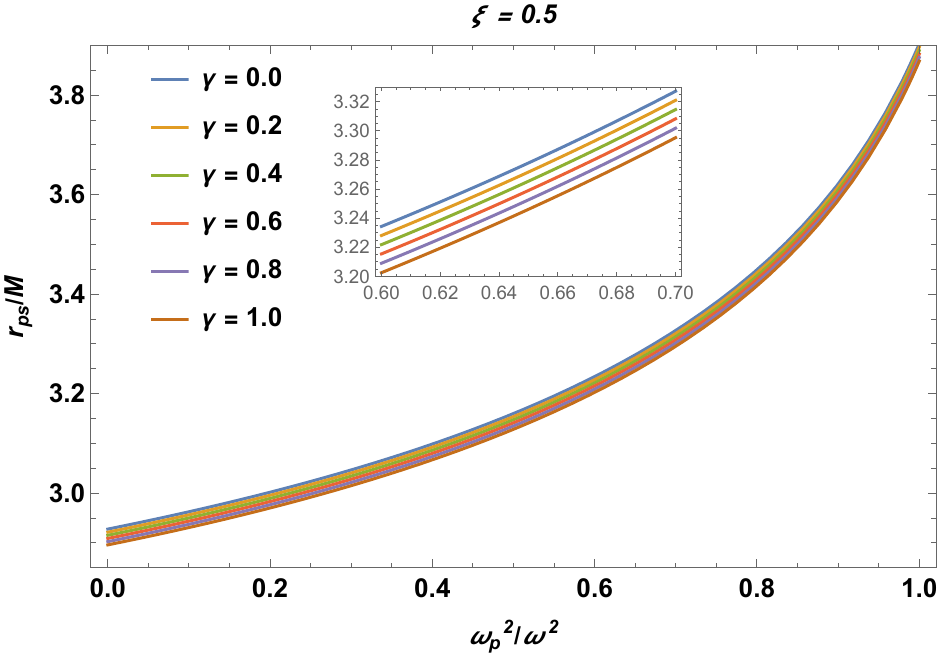}
  \end{center}
\caption{{The plot demonstrates the photon sphere radii as a function of the plasma frequency for the different values of the $\xi$ (left panel) and $\gamma$ (right panel) parameters. Here, we set $\gamma=0.5$ and $\xi=0.5$ for the left and right panels, respectively.}}
\label{fig:photonradiusuni}
\end{figure*}
\begin{figure*}[!htb]
    \centering
     \includegraphics[scale=0.5]{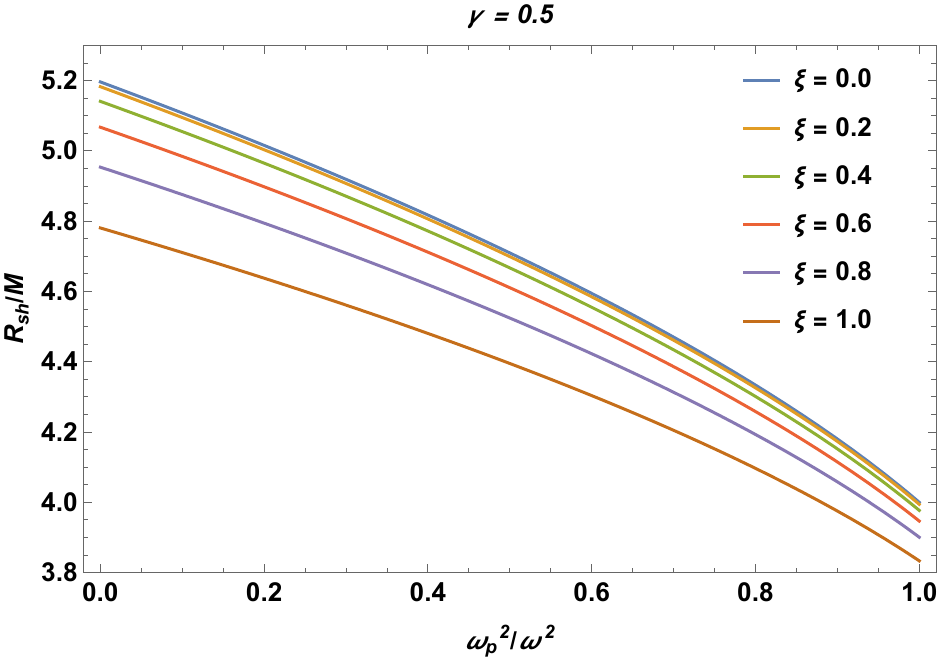}
      \includegraphics[scale=0.5]{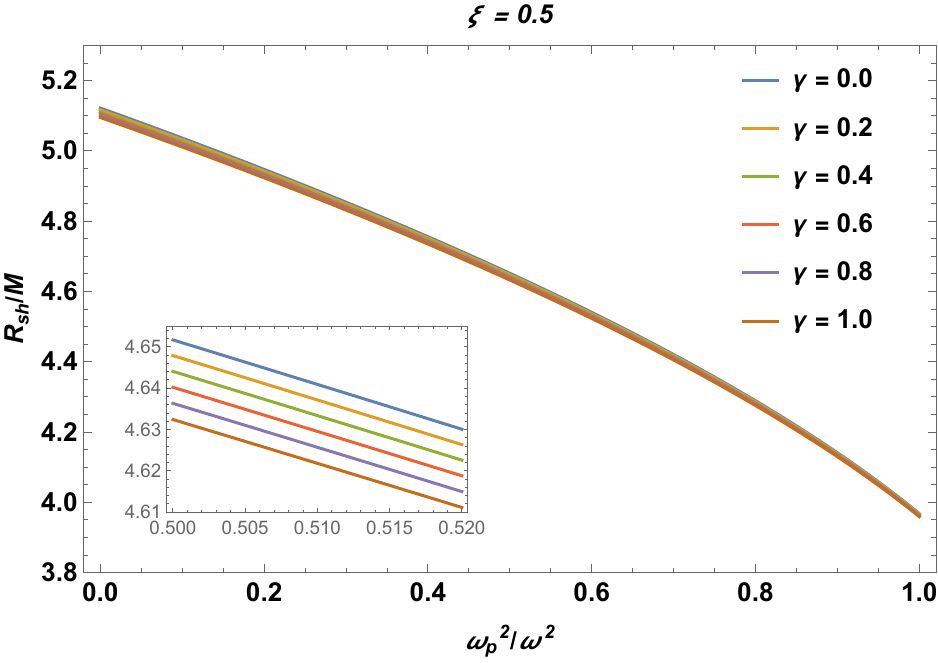}
    \caption{{The plot shows the dependence of the BH shadow radii on the plasma frequency for the different values of the $\xi$ (left panel) and $\gamma$ (right panel) parameters. Here, we set $\gamma=0.5$ and $\xi=0.5$ for the left and right panels, respectively.} }
    \label{shadow1}
\end{figure*}
\begin{figure*}[!htb]
    \centering
     \includegraphics[scale=0.55]{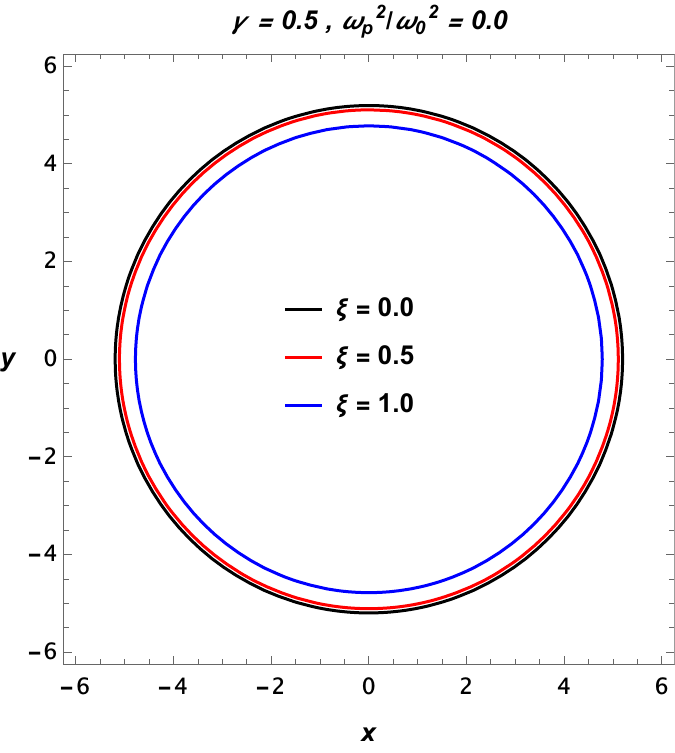}
      \includegraphics[scale=0.55]{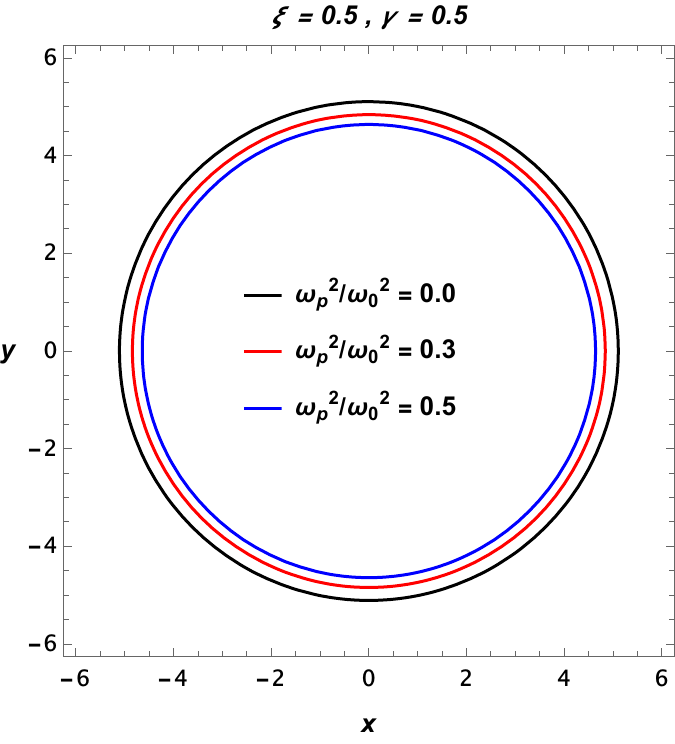}
    \caption{{The plot illustrates the profile of the shadow cast by the Schwarzschild-like BH for the different values of the $\xi$ parameter and the plasma frequency.}}
    \label{shadow2}
\end{figure*}
\begin{figure*}[!htb]
    \centering
    \includegraphics[scale=0.55]{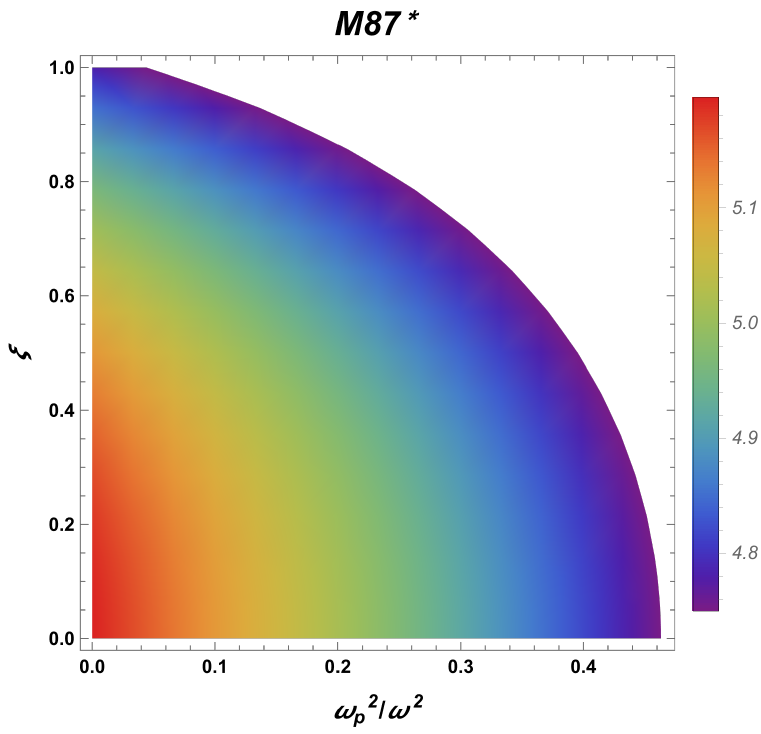}
    \includegraphics[scale=0.55]{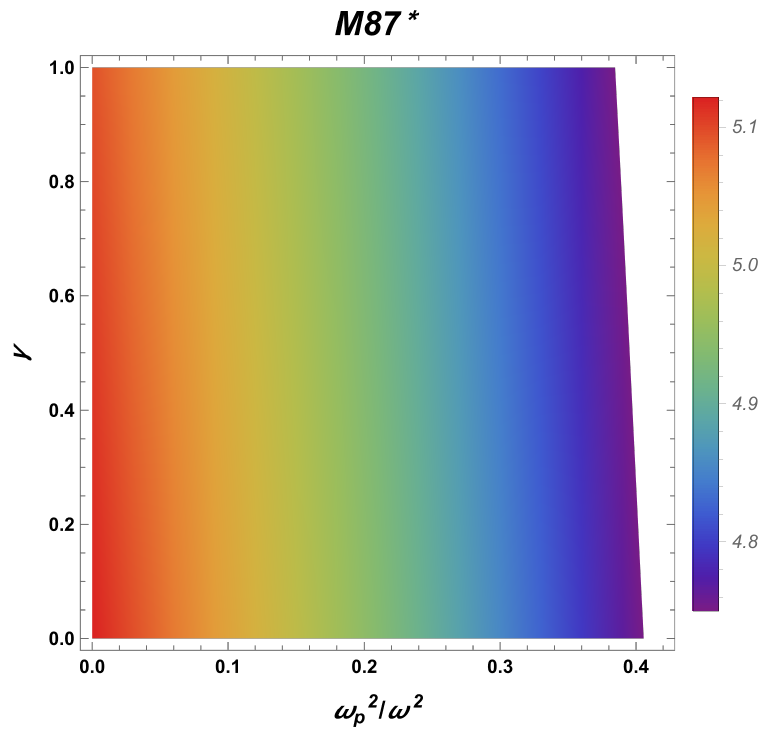}
    \includegraphics[scale=0.55]{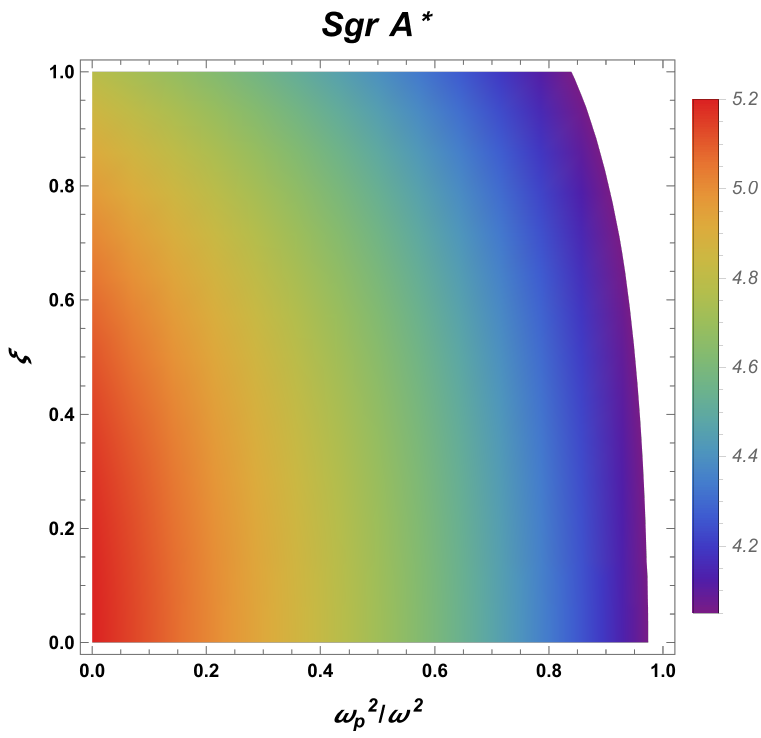}
    \includegraphics[scale=0.55]{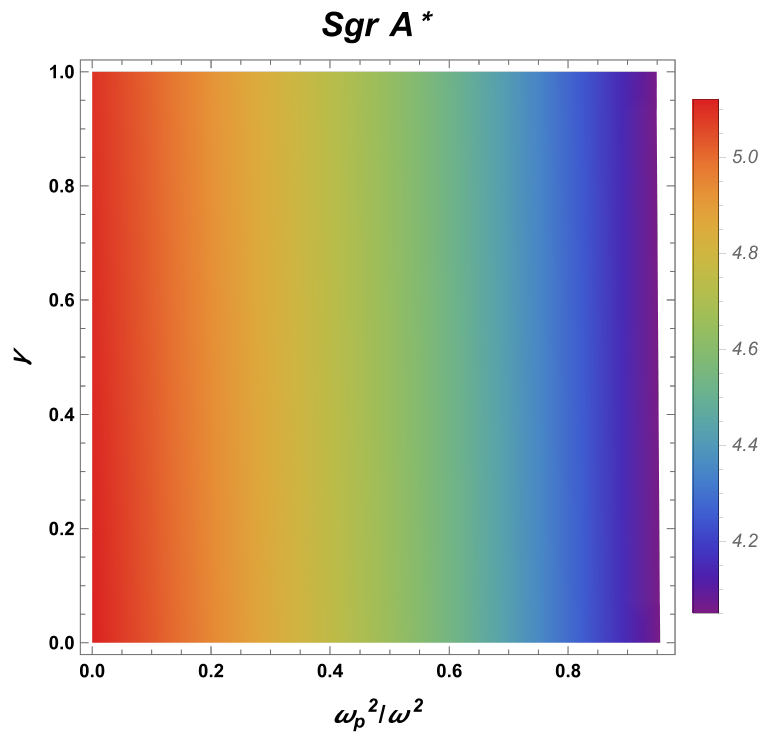}
    \caption{{The plot demonstrated the constrained values of spacetime parameters and the plasma frequency for the M87* (top panels) and Sgr A* (bottom panels). Here, we set $\gamma=0.5$ and $\xi=0.5$ for the left panels and right panels, respectively.}}
    \label{fig:constrain}
\end{figure*}

Now, we investigate the null geodesics around the Schwarzschild-like BH in the presence of the plasma by using the Hamilton-Jacobi equation. The Hamiltonian for the null geodesics around BH in the presence of the plasma can be written in the following form~\cite{Synge:1960b}
\begin{equation}
\mathcal H(x^\alpha, p_\alpha)=\frac{1}{2}\left[ g^{\alpha \beta} p_\alpha p_\beta - (n^2-1)( p_\beta u^\beta )^2 \right]\ ,
\label{gH}
\end{equation}
where $x^\alpha$, $u^\beta$, and $p_\alpha$, respectively, refer to the spacetime coordinate, four-velocity, and photon's momentum. Also, $n$ represents the refractive index, and it can be defined as~\cite{Tsp:2009a}
\begin{eqnarray}
n^2&=&1- \frac{\omega_{\text{p}}^2}{\omega^2}\, ,
\label{eq:n1}
\end{eqnarray}
where $\omega_p$ and $\omega$ are the frequencies of plasma and photon, respectively. One can define them as
\begin{eqnarray}
&&\omega^2_{p}(x^\alpha)=4 \pi e^2 N(x^\alpha)/m_e\,,\nonumber\\
&&\omega(r)=\frac{\omega_0}{\sqrt{f(r)}}\ ,\quad  \omega_0=\text{const}\,,
\end{eqnarray}
with $m_e$, $e$ and $N$, respectively, represent electron mass, electron charge and number density of electrons. The metric function satisfies $f(r) \rightarrow1$ as $r\rightarrow\infty$, while $\omega(\infty)=\omega_0=-p_t$ defines the energy of a photon at spatial infinity. $\omega_0$ can be restricted by using the ${\cal H}=0$ as
\begin{equation}
 \frac{\omega^2_0}{f(r)}>\omega^2_p(r)\, .   
\end{equation}
This condition physically implies that the photon's local frequency, $\omega(r)$, must exceed the plasma frequency at that location. Because this rule governs light in plasma, the resulting BH shadow can exhibit different forms than it does in a vacuum case ($\omega_p=0$). 
Using the Eq.~(\ref{eq:n1}), one can rewrite the Eq.~(\ref{gH}) as follows~\cite{Synge:1960b,Rog:2015a}
\begin{equation}
\mathcal{H}=\frac{1}{2}\Big[g^{\alpha\beta}p_{\alpha}p_{\beta}+\omega^2_{\text{p}}]\, . 
\end{equation}
After that, the light ray equations for the photon can be obtained by using $\dot x^\alpha=\partial \mathcal{H}/\partial p_\alpha$ in the equatorial plane ($\theta=\pi/2$) as~\cite{Chandrasekhar1983}
\begin{eqnarray} 
\dot t\equiv\frac{dt}{d\lambda}&=& -\frac{ {p_t}}{f(r)}\, ,  \\
\dot r\equiv\frac{dr}{d\lambda}&=&p_r f(r)\, , \label{eqr1} \\
\dot\phi\equiv\frac{d \phi}{d\lambda}&=&  \frac{p_{\phi}}{r^2}\, , \label{eqvarphi1}
\end{eqnarray}
We then get the orbit equation by taking the ratio of the Eqs.~(\ref{eqr1}) and (\ref{eqvarphi1}) as
\begin{equation}
\frac{dr}{d\phi}=\frac{g^{rr}p_r}{g^{\phi\phi}p_{\phi}}\, .    \label{trajectory}
\end{equation}
The above equation can be rewritten for the photon geodesics $\mathcal H=0$ in the following form~\cite{Babar21a}
\begin{equation}
 \frac{dr}{d\phi}=\sqrt{\frac{g^{rr}}{g^{\phi\phi}}}\sqrt{h^2(r)\frac{\omega^2_0}{p_\phi^2}-1}\, ,
\end{equation}
with
\begin{equation}
    h^2(r)\equiv-\frac{g^{tt}}{g^{\phi\phi}}-\frac{\omega^2_p}{g^{\phi\phi}\omega^2_0}\, . 
\end{equation}
When a light ray travels from infinity, reaches its closest approach at radius $r_{ps}$, and returns to infinity, this trajectory corresponds to a turning point in the $\gamma^2(r)$ function. Consequently, the photon sphere radius is obtained by solving the equation below.
\begin{equation}
\frac{d(h^2(r))}{dr}\bigg|_{r=r_{\text{ps}}}=0\, .     
\end{equation}
We numerically investigate the photon sphere radii due to the complex spacetime of the Schwarzschild-like BH. The Fig.~\ref{fig:photonradiusuni} shows the photon sphere radii as a function of plasma frequency for the different values of the $\xi$ and $\gamma$ parameters. As seen from this figure, the values of the radius of the photon sphere increase with the rise of the plasma frequency and vice versa for the $\xi$ parameter. There is a slight decrease in the photon sphere radii under the influence of the $\gamma$ parameter.

\subsection{Black hole shadow in the presence of plasma}

{
This subsection is devoted to the investigation of the BH shadow in plasma. One can write the angular radius of the shadow~\cite{Perlick15a,Konoplya2019} 
\begin{eqnarray}
\sin^2 \alpha_{\text{sh}}&=&\frac{h^2(r_{\text{ps}})}{h^2(r_{\text{o}})}=\frac{r^2_{\text{ps}}\left[\frac{1}{f(r_{\text{ps}})}-\frac{\omega^2_p(r_{\text{ps}})}{\omega^2_0}\right]}{r^2_{\text{o}}\left[\frac{1}{f(r_{\text{o}})}-\frac{\omega^2_p(r_{\text{o}})}{\omega^2_0}\right]}\,,
\end{eqnarray}
with $r_{\text{o}}$ and $r_{\text{ps}}$ referring to the locations of the observer and the photon sphere, respectively. We can approximate the BH shadow in the following form by assuming that the observer is located at a large distance from the BH~\cite{Perlick15a}
\begin{eqnarray}
R_{\text{sh}}&\simeq& r_{\text{o}} \sin \alpha_{\text{sh}},\\
 &=&\sqrt{r^2_{\text{ps}}\left[\frac{1}{f(r_{\text{ps}})}-\frac{\omega^2_p(r_{\text{ps}})}{\omega^2_0}\right]}\,.  \nonumber
\end{eqnarray}
In the Fig.~\ref{shadow1}, we demonstrate the shadow radii as a function of the plasma frequency for the different values of the $\xi$ and $\gamma$ parameters. It can be observed from this figure that the radius of the BH shadow decreases with the increase of both plasma frequency and $\xi$ parameter. There is also a slight decrease under the influence of the $\gamma$ parameter. To provide more information, we explore the appearance of the BH shadow from the perspective of a distant observer. Therefore, we can write the following expressions by using the celestial coordinates~\cite{vazquez_esteban_2004}
\begin{eqnarray}
X&=&\lim_{r_{0}\rightarrow \infty }\left( -r_{0}\sin \theta _{0}\left. \frac{
d\varphi }{dr}\right\vert _{r_{0},\theta _{0}}\right)  \,,  \label{X1}\\
Y&=&\lim_{r_{0}\rightarrow \infty }\left( r_{0}\left. \frac{d\theta }{dr}
\right\vert _{r_{0},\theta _{0}}\right) ,  \label{Y1}
\end{eqnarray}
where $(r_0,\theta_0)$ represent the position of observer. The Eqs.~(\ref{X1}) and (\ref{Y1}) obey the following relation if we assume that the observer is located on the equatorial hyperplane
\begin{equation}
X^{2}+Y^{2}=R_{sh}^{2}\,.
\end{equation} 
Using the above equation, we plot the BH shadow in Fig.~\ref{shadow2}. It can be seen from this figure that the BH shadow shrinks with the increase of the $\xi$ parameter and the plasma frequency. 
\begin{table}[htbp]
\centering
\caption{Observational data for M87* and Sgr A*~\cite{EventHorizonTelescope:2024dhe,DeLaurentis:2022nrv,GRAVITY:2020gka}.}
\label{tabdata1}
\resizebox{0.5\textwidth}{!}{
\begin{tabular}{|l|c|c|}
\hline
\textbf{Parameter}         & \textbf{M87*} & \textbf{Sgr A*} \\ \hline
Angular Diameter ($\theta$) & $43.3 \pm 2.3 \, \mu\text{as}$ & $51.8 \pm 2.3 \, \mu\text{as}$ \\ \hline
Distance ($D$)             & $16.5 \, \text{Mpc}$ & ${8.275 \, \text{kpc}}$ \\ \hline
Mass ($M$)                 & $(6.5 \pm 0.7) \times 10^9 \, M_{\odot}$ & $(4.297 \pm 0.013) \times 10^6 \, M_{\odot}$ \\ \hline
\end{tabular}
}
\end{table}
Moreover, we get the theoretical constraints for the parameters of the spacetime and the plasma frequency by using the observational data of the EHT and GRAVITY collaborations with the assumption that M87* and Sgr A* are static and spherically symmetric, inspite of the observations that do not support it. To do so, we can use the following equation to calculate the shadow diameter~\cite{Bambi_2019}
\begin{equation}
    D_{sh}=\frac{D\theta}{M}\, .
\end{equation}
Using the observational data in Table~\ref{tabdata1}, one can write the diameter of the shadow as
\begin{eqnarray}
D^{\text{M87}^*}_{sh}&=&(11 \pm 1.5)M \,,\nonumber\\
D^{\text{Sgr A}^*}_{sh}&=&(9.5 \pm 1.4)M\,.
\end{eqnarray}
Using the simple equation $D_{sh}=2R_{sh}$, one can constrain the values of the spacetime parameters and the plasma frequency. These results were demonstrated in the top panel (for M87*) and the bottom panel (for Sgr A*) of Fig.~\ref{fig:constrain}.  
}

\section{Weak gravitational lensing for black hole}\label{3}

{This part is devoted to the framework of the gravitational weak lensing around the Schwarzschild-like BH in the presence of the uniform and non-uniform plasma cases. Therefore, we need to define the weak-field approximation as~\cite{BisnovatyiKogan2010,Babar21a}
\begin{equation}
    g_{\alpha \beta}=\eta_{\alpha \beta}+h_{\alpha \beta}\, .
\end{equation}
Here, $\eta_{\alpha \beta}$ and $h_{\alpha \beta}$ represent the Minkowski spacetime and the perturbation gravity, respectively. Therefore, the following conditions are suitable for them
\begin{eqnarray}
 \eta_{\alpha \beta}&=&diag(-1,1,1,1)\ , \nonumber\\
 h_{\alpha \beta} &\ll& 1, \hspace{0.5cm} h_{\alpha \beta} \rightarrow 0 \hspace{0.5cm} \mbox{under} \hspace{0.2cm}  x^{\alpha}\rightarrow \infty \ ,\nonumber\\
 g^{\alpha \beta}&=&\eta^{\alpha \beta}-h^{\alpha \beta}, \hspace{0,5cm} h^{\alpha \beta}=h_{\alpha \beta}\ \,.
\end{eqnarray}
One can use the following equation to investigate the deflection angle around the BH~\cite{BisnovatyiKogan2010}
\begin{eqnarray}
    \hat{\alpha }_{\text{b}}&=&\frac{1}{2}\int_{-\infty}^{\infty}\frac{b}{r}\left(\frac{dh_{33}}{dr}+\frac{1}{1-\omega^2_p/ \omega^2}\frac{dh_{00}}{dr}-\right. \nonumber\\
    &-&\left.\frac{K_e}{\omega^2-\omega^2_p}\frac{dN}{dr} \right)dz\, , 
\end{eqnarray} 
with $\omega$ and $\omega_{p}$ are respectively referring to the frequencies of photon and plasma. To explore the weak-gravitational lensing, we need to expand the line element of the Schwarzschild-like BH into a Taylor series in the following form
\begin{eqnarray}
  ds^2 &\approx& ds^2_0+\frac{4Mr^2}{\xi^2(\gamma M+r)+\sqrt{\xi^4(\gamma M+r)^2+4r^6}} dt^2 +\nonumber\\
  &+&\frac{4Mr^2}{\xi^2(\gamma M+r)+\sqrt{\xi^4(\gamma M+r)^2+4r^6}}dr^2\, .
\end{eqnarray}
where $ds^2_0=-dt^2+dr^2+r^2(d\theta^2+\sin^2\theta d\phi^2)$. We further write the components of $h_{\alpha\beta}$ as
\begin{eqnarray}
     h_{00}&=&\frac{4Mr^2}{\xi^2(\gamma M+r)+\sqrt{\xi^4(\gamma M+r)^2+4r^6}}\, ,\nonumber\\
    h_{ik}&=&\frac{4Mr^2}{\xi^2(\gamma M+r)+\sqrt{\xi^4(\gamma M+r)^2+4r^6}}n_i n_k\, , \nonumber\\
    h_{33}&=&\frac{4Mr^2}{\xi^2(\gamma M+r)+\sqrt{\xi^4(\gamma M+r)^2+4r^6}} \cos^2\chi \label{dh}\, ,
\end{eqnarray}
where $\cos^2\chi=z^2/(b^2+z^2)$ and $r^2=b^2+z^2$. After that, it is possible to write the deflection angle as a combination of three angles as follows~\cite{Atamurotov2021}
\begin{eqnarray}\label{ab}
\hat{\alpha_b}=\hat{\alpha_1}+\hat{\alpha_2}+\hat{\alpha_3}\, , \label{ae}
\end{eqnarray}
with  
 \begin{eqnarray}
\hat{\alpha_1}&=&\frac{1}{2}\int_{-\infty}^{\infty} \frac{b}{r}\frac{dh_{33}}{dr}dz\ ,\nonumber\\
\hat{\alpha_2}&=&\frac{1}{2}\int_{-\infty}^{\infty} \frac{b}{r}\frac{1}{1-\omega^2_p/ \omega^2}\frac{dh_{00}}{dr}dz\ ,\nonumber\\
\hat{\alpha_3}&=&\frac{1}{2}\int_{-\infty}^{\infty} \frac{b}{r}\left(-\frac{K_e}{\omega^2-\omega^2_p}\frac{dN}{dr} \right)dz\ .  \label{ma}
\end{eqnarray}
In the subsequent parts, we investigate the deflection angle in the presence of the uniform and non-uniform plasma density distributions. 
}

\begin{figure*}[!htb]
    \centering
     \includegraphics[width=0.4\linewidth]{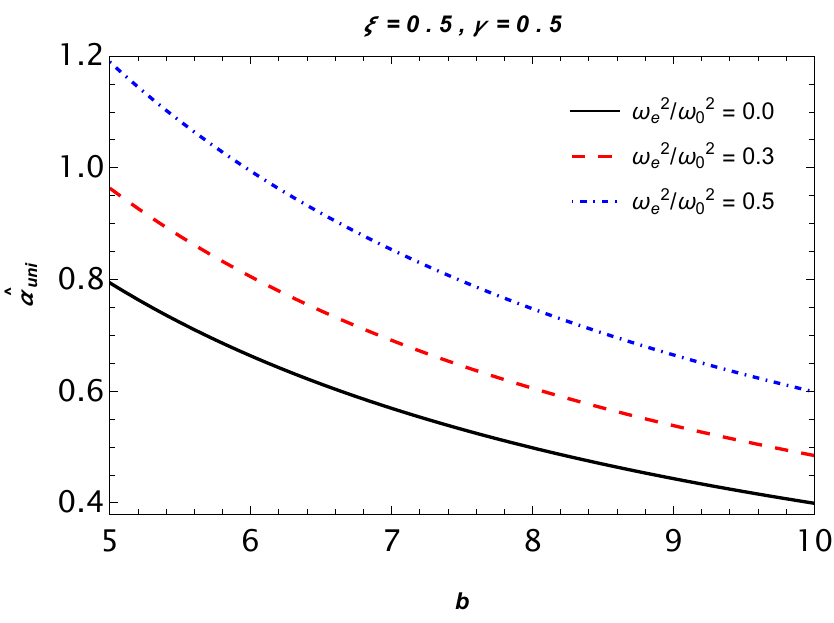}
    \includegraphics[width=0.41\linewidth]{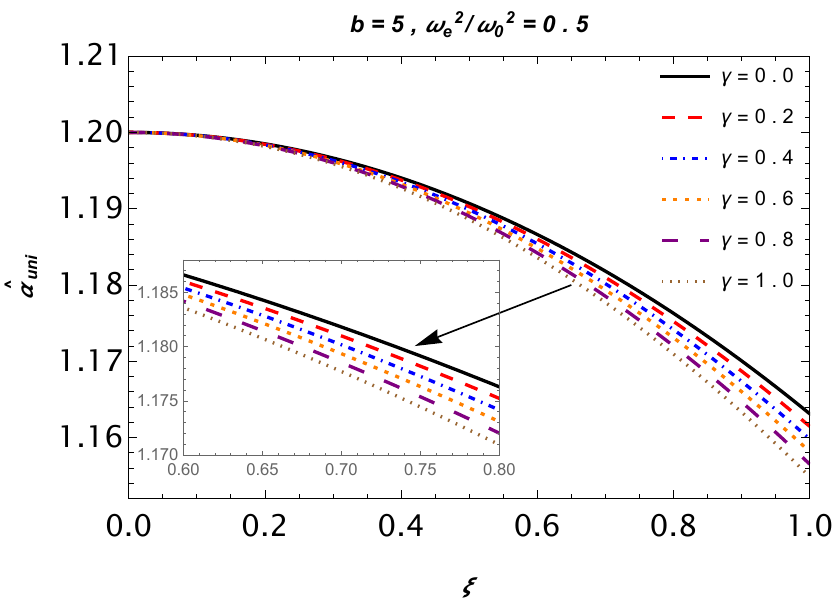}  
     \caption{{The left panel shows the deflection angle as a function of the impact parameter for the different values of the uniform plasma frequency. Here, we set the spacetime parameters as $\xi=0.5$ and $\gamma=0.5$. Right panel: the dependence of the deflection angle on the $\xi$ parameter for the different values of the $\gamma$ parameter. The impact parameter and the plasma frequency were fixed for this panel, i.e., $b=5$ and $\omega_e^2/\omega_0^2=0.5$.}}
    \label{fig:uni}
\end{figure*}

\subsection{Uniform plasma}

{The Eq.~(\ref{ab}) can be written for the uniform plasma case as~\cite{Tsp:2015a,Atamurotov2021}
\begin{equation}
\hat{\alpha}_{uni}=\hat{\alpha}_{uni1}+\hat{\alpha}_{uni2}+\hat{\alpha}_{uni3}\, .
\end{equation}
We can analyze the deflection angle numerically by combining the Eqs.~(\ref{dh}), (\ref{ae}) and~(\ref{ma}). The left panel of Fig.~\ref{fig:uni} demonstrates the deflection angle as a function of the impact parameter for the different values of the plasma frequency. As shown in this figure, the values of the deflection angle decrease with the rise of the impact parameter and vice versa for the uniform plasma frequency. Here, we set the spacetime parameters as $\xi=\gamma=0.5$. We plot the dependence of the $\hat{\alpha}_{uni}$ on the $\xi$ parameter for the different values of the $\gamma$ parameter in the right panel of Fig.~\ref{fig:uni}. One can observe from this panel that the deflection angle decreases with increasing values of both $\xi$ and $\gamma$. 
}

\subsection{Non uniform plasma}

\begin{figure*}
    \centering
     \includegraphics[width=0.4\linewidth]{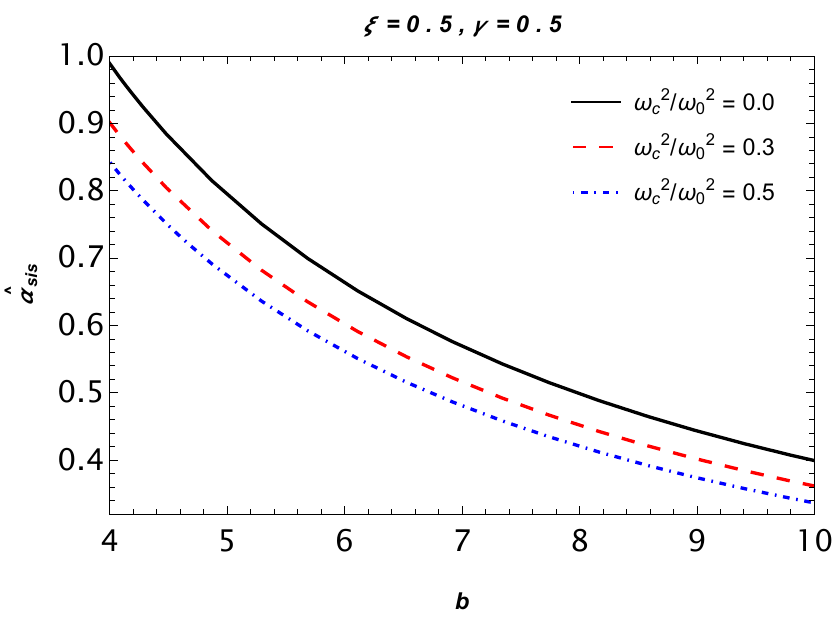}
    \includegraphics[width=0.42\linewidth]{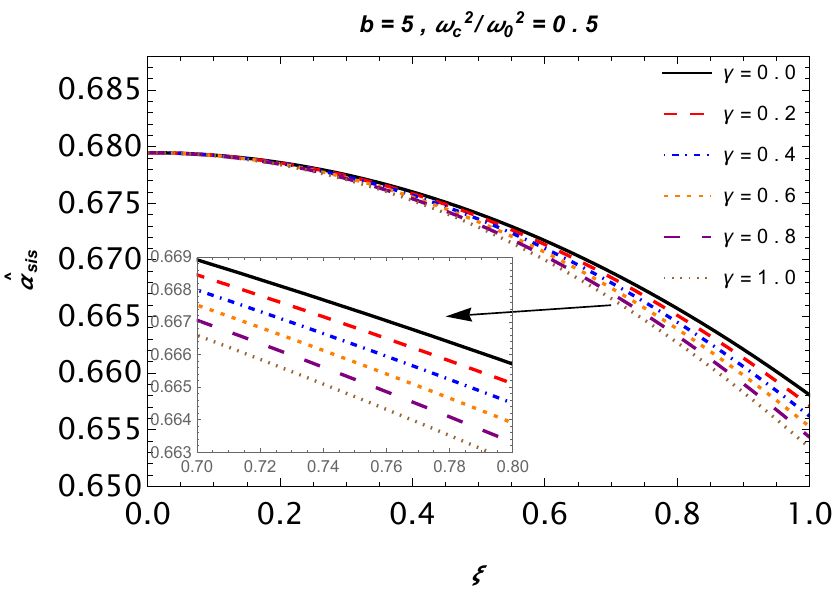}  
    \caption{{The same as Fig.~\ref{fig:uni} but for the non-uniform plasma distribution.}}
    \label{fig:sis}
\end{figure*}

\begin{figure*}
    \centering
    \includegraphics[width=0.45\linewidth]{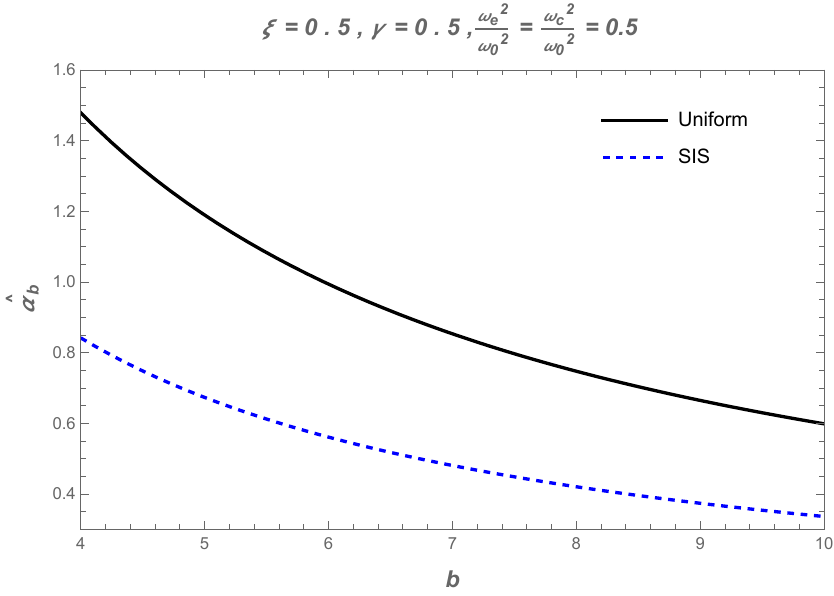}
    \includegraphics[width=0.45\linewidth]{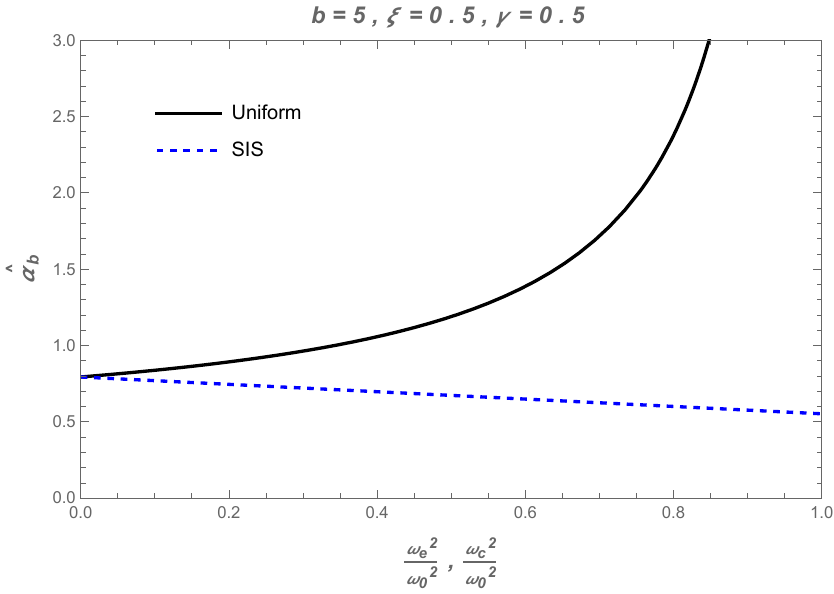}
    \caption{{Illustrated in the plot is a comparison of the deflection angle between uniform and non-uniform plasma distributions. The dependence on the impact parameter is shown in the left panel, and the dependence on the plasma frequency is shown in the right panel.}}
    \label{fig:compare}
\end{figure*}

{In this subsection, we examine the singular isothermal sphere (SIS), which stands as the most widely favored model for probing photon geodesics near BHs. Typically described as a spherical gas cloud, it features a central singularity where density becomes theoretically infinite. The SIS density distribution can be written as follows~\cite{BisnovatyiKogan2010}
\begin{equation}
    \rho(r)=\frac{\sigma^2_{\nu}}{2\pi r^2}\, ,
\end{equation}
with $\sigma^2_{\nu}$ is referring an one-dimensional velocity dispersion. Subsequently, we can write the plasma frequency in the following form
\begin{equation}
    \omega^2_c=K_e N(r)=\frac{K_e \sigma^2_{\nu}}{2\pi k m_p r^2}\, .
\end{equation}
Keeping this in mind, the deflection angle for the non-uniform plasma case can be written by combining three angles as
\begin{equation}
\hat{\alpha}_{SIS}=\hat{\alpha}_{SIS1}+\hat{\alpha}_{SIS2}+\hat{\alpha}_{SIS3} \, .
\end{equation}
We plot the dependence of the deflection angle in the presence of the non-uniform plasma on the impact parameter for the different values of the plasma frequency in the left panel of Fig.~\ref{fig:sis}. It can be seen from this panel that there is a decrease with the increase in the values of the impact parameter and plasma frequency. Furthermore, the deflection angle for the non-uniform plasma was plotted as a function of the $\xi$ parameter for the different values of the $\gamma$ parameter. As shown in this panel, there is a slight decrease under the influence of both $\xi$ and $\gamma$ parameters. It should be noted that the effects of uniform and non-uniform plasma cases on the deflection angle are opposite to each other. 
We compare the deflection angle of light for the uniform and non-uniform plasma cases in Fig.~\ref{fig:compare} to provide more information. 
}


\section{Magnification of gravitationally lensed image}\label{4}

This section is devoted to the magnification of the gravitationally lensed image around the Schwarzschild-like BH surrounded by the uniform and non-uniform plasma.  One can define the magnification factors in the presence of the plasma, which are $\mu_{+}^{pl}$ and $\mu_-^{pl}$, referring to the primary and secondary images, respectively, as~\cite{Schneider1992,Atamurotov:2021cgh,Alloqulov_2023,Alloqulov_2024}
\begin{eqnarray}
\mu^\mathrm{pl}_\mathrm{+}=\frac{1}{4}\bigg(\frac{x}{\sqrt{x^2+4}}+\frac{\sqrt{x^2+4}}{x}+2\bigg)\, ,
\end{eqnarray}
\begin{eqnarray}
\mu^\mathrm{pl}_\mathrm{-}=\frac{1}{4}\bigg(\frac{x}{\sqrt{x^2+4}}+\frac{\sqrt{x^2+4}}{x}-2\bigg)\, ,
\end{eqnarray}
where~\cite{Babar21a}
\begin{eqnarray}
x&=&{\beta}/{\theta_E}\,,
\end{eqnarray}
with $\theta_E$ referring to Einstein's ring, which defines the shape of the image.  After that, one can get the total magnification as a combination of the magnification factors as~\cite{Tsp:2015a}
\begin{eqnarray}\label{totmag}
\mu^\mathrm{pl}_\mathrm{tot}=\mu^\mathrm{pl}_{+}+\mu^\mathrm{pl}_{-}=\frac{x^2+2}{x\sqrt{x^2+4}}\, .
\end{eqnarray}
In the next subsections, we analyze the effect of the uniform and non-uniform plasma on the total magnification.

\subsection{Uniform Plasma}

\begin{figure*}[!htb]
    \centering
    \includegraphics[scale=0.5]{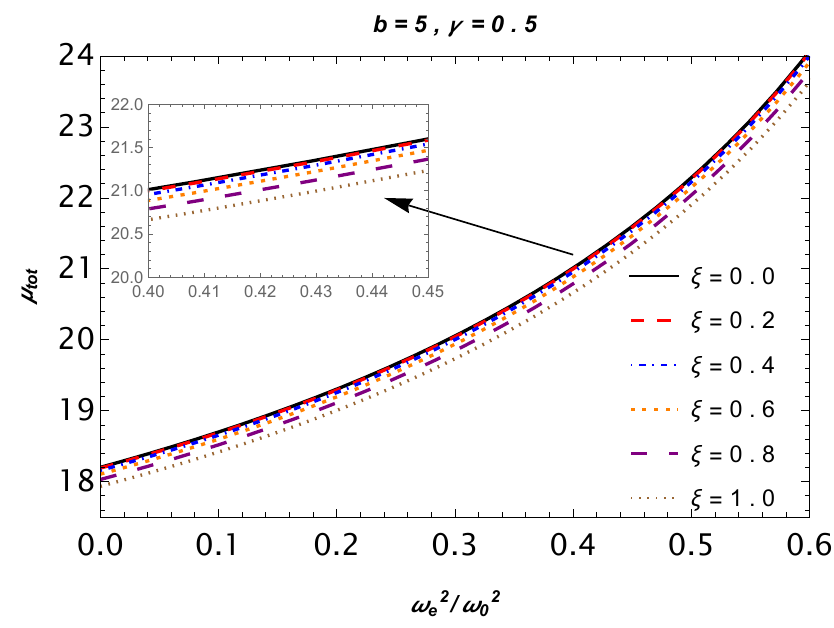}
    \includegraphics[scale=0.5]{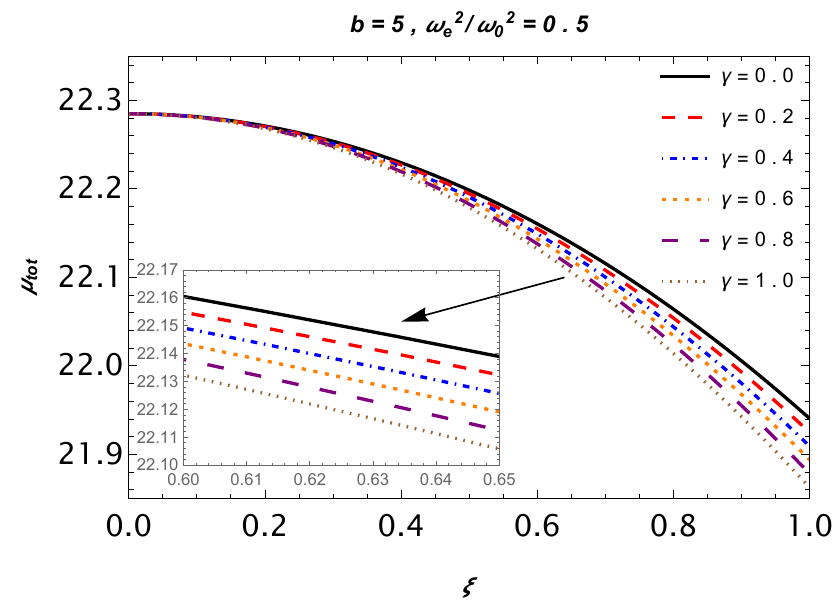}
    \caption{{Left panel: the total magnification as a function of the uniform plasma frequency for the different values of the $\xi$ parameter, while other parameters were fixed as $b=5$ and $\gamma=0.5$. The right panel shows the dependence of the total magnification on the $\xi$ parameter for the different values of the $\gamma$ parameter. Here, we set $b=5$ and $\omega_e^2/\omega_0^2=0.5$. } }
    \label{magf1}
\end{figure*}
\begin{figure*}[!htb]
    \centering
    \includegraphics[scale=0.5]{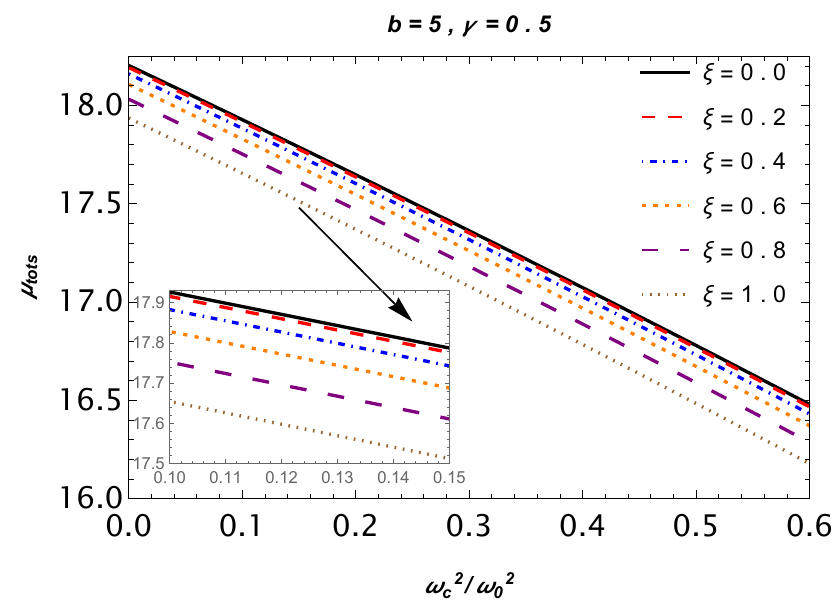}
    \includegraphics[scale=0.5]{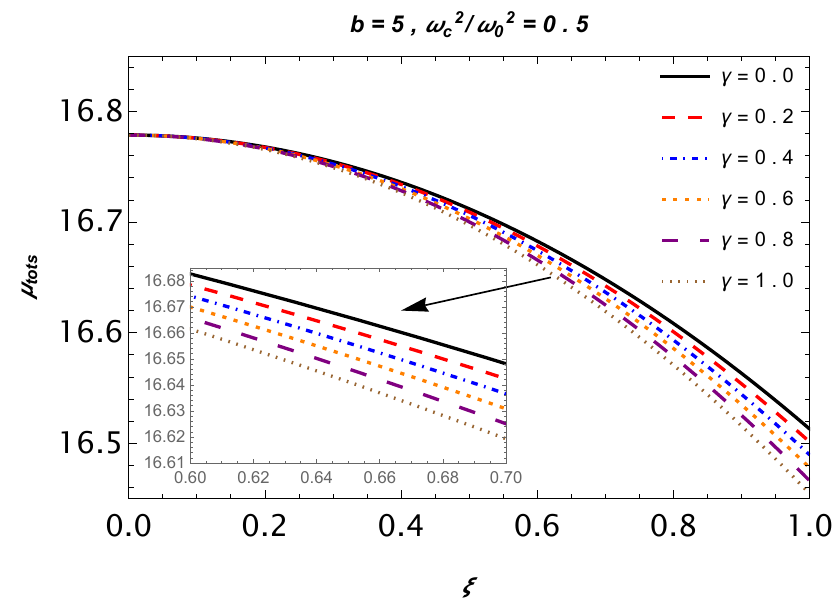}
    \caption{{The same as Fig.~\ref{magf1} but for the non-uniform plasma scenario.}}
    \label{magf2}
\end{figure*}
\begin{figure*}[!htb]
    \centering
    \includegraphics[width=0.4\linewidth]{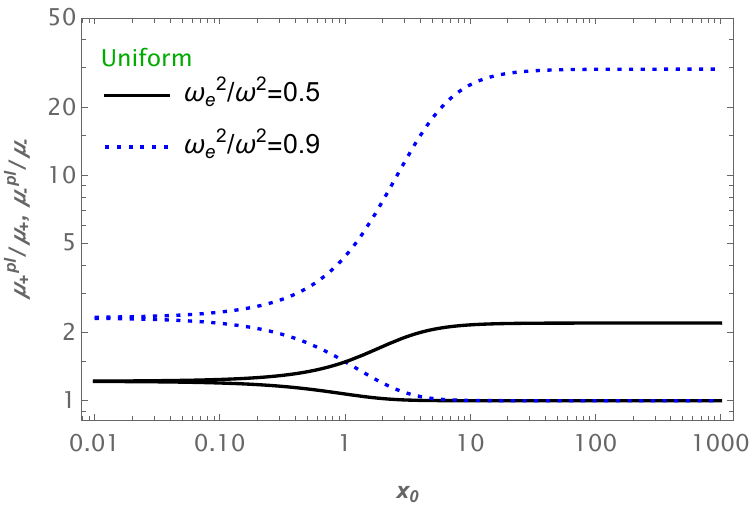}
    \includegraphics[width=0.4\linewidth]{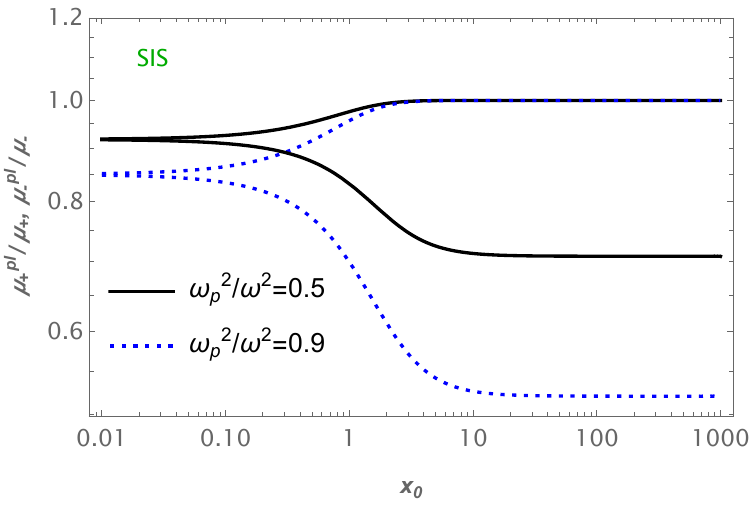}
    \includegraphics[width=0.4\linewidth]{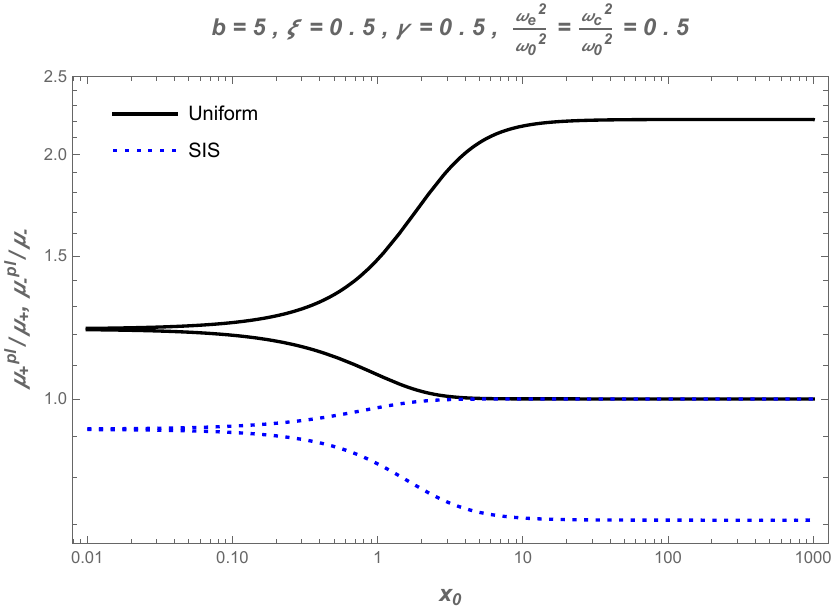}
     \includegraphics[width=0.4\linewidth]{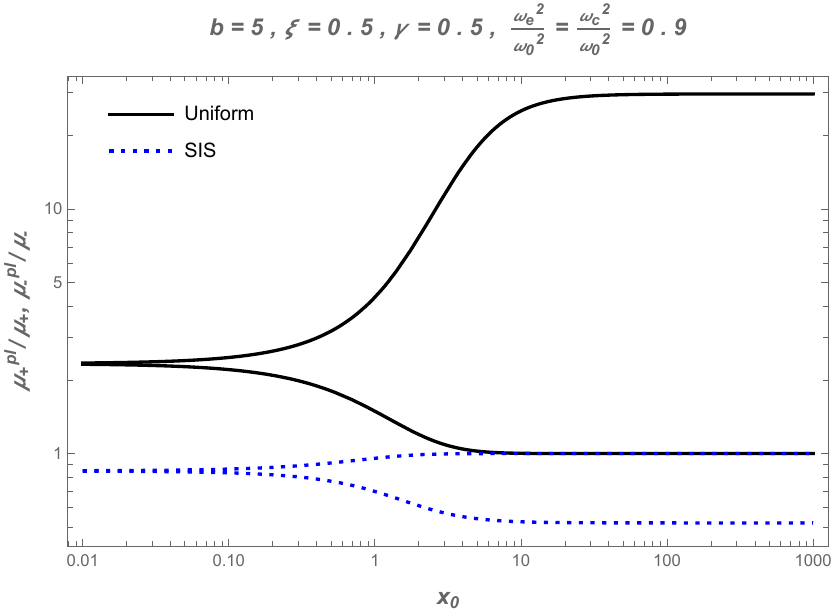}
    \caption{{The top left and right panels illustrate the ratio of the magnification factors in the presence of plasma to that of the vacuum case as a function of $x_0$ for the different values of the plasma frequencies, while their comparisons were presented in the bottom panels.}}
    \label{magf3}
\end{figure*}

In this part, we consider the uniform plasma. Therefore, the Eq.~(\ref{totmag}) can be rewritten as follows
\begin{equation}
\Big(\mu^{pl}_{tot}\Big)_{uni}=\Big(\mu^{pl}_{+}\Big)_{uni}+\Big(\mu^{pl}_{-}\Big)_{uni}=\dfrac{x^2_{uni}+2}{x_{uni}\sqrt{x^2_{uni}+4}}\, ,
\end{equation}
where 
\begin{eqnarray}
       (\mu^{pl}_{\pm})_{uni}&=&\frac{1}{4}\left(\dfrac{x_{uni}}{\sqrt{x^2_{uni}+4}}+\dfrac{\sqrt{x^2_{uni}+4}}{x_{uni}}\pm2\right)\, ,\nonumber \\
       x_{uni}&=&\frac{\beta}{(\theta^{pl}_E)_{uni}}\, .
\end{eqnarray}
We plot the total magnification as a function of the uniform plasma frequency for the different values of the $\xi$ parameter while other parameters are fixed as $b=5 , \quad \gamma=0.5$ in the left panel of Fig.~\ref{magf1}. One can see from this panel that the total magnification increases with the rise in the uniform plasma frequency. Also, there is a slight decrease under the influence of the $\xi$ parameter. In the right panel of this figure, the dependence of the total magnification on the $\xi$ parameter for the different values of the $\gamma$ parameter was plotted. It is clear from this panel that the total magnification decreases due to the rise of the spacetime parameters. Here, the other parameters were considered as a constant, as $b=5$ and $\omega_e^2/\omega_0^2=0.5$.

\subsection{Non-uniform plasma}

{As performed in the previous part, we examine the effect of the non-uniform plasma on the total magnification. Hence, we recall the Eq.~(\ref{totmag}) for the non-uniform plasma (SIS medium)  as~\cite{BisnovatyiKogan2010}
\begin{equation}
    \Big(\mu^{pl}_{tot}\Big)_{SIS}=\Big(\mu^{pl}_{+}\Big)_{SIS}+\Big(\mu^{pl}_{-}\Big)_{SIS}=\dfrac{x^2_{SIS}+2}{x_{SIS}\sqrt{x^2_{SIS}+4}}\, ,
\end{equation}
with
\begin{eqnarray}
       (\mu^{pl}_{\pm})_{SIS}&=&\frac{1}{4}\left(\dfrac{x_{SIS}}{\sqrt{x^2_{SIS}+4}}+\dfrac{\sqrt{x^2_{SIS}+4}}{x_{SIS}}\pm2\right)\ \,, \nonumber \\
        x_{SIS}&=&\frac{\beta}{(\theta^{pl}_E)_{SIS}}\, .
\end{eqnarray}
Using the above equations, we plot the dependence of the total magnification on the non-uniform plasma frequency for the different values of the $\xi$ parameter in the left panel of Fig.~\ref{magf2}. We can see from this panel that the total magnification decreases under the influence of both plasma frequency and $\xi$ parameter. Furthermore, the total magnification as a function of the $\xi$ parameter was plotted for the different values of the $\gamma$ parameter in the right panel of Fig.~\ref{magf2}. It can be seen from this panel that there is a slight decrease with the increase of the spacetime parameters. Additionally, we plot the ratio of the magnification factors to that of in vacuum case as a function of the $x_0$ for the uniform and non-uniform plasma cases together with the comparison of these cases in Fig.~\ref{magf3}. Notably, this ratio increases with the increase of the plasma frequency in the uniform plasma case and vice versa for the non-uniform plasma. Also, the uniform plasma distribution exhibits greater sensitivity to this ratio compared to the non-uniform one. 
}

\section{Conclusions}\label{con}

In this work, we analyze the impact of the uniform and non-uniform plasma on the shadow and weak gravitational lensing around the Schwarzschild-like BH. One can summarize our key results based on the performed research in the following way.

First, we perform an analysis of the spacetime geometry by plotting the radial dependence of the metric function for the different values of the $\xi$ and $\gamma$ parameters. We then turn to investigate the event horizon structure of the Schwarzschild-like BH. It was found that the event horizon radii decrease due to the rise of both spacetime parameters. In addition, using the Hamiltonian formalism, we study the photon dynamics around the BH in the presence of the homogeneous plasma together with the BH shadow. The outcomes indicate that the photon sphere radii increase under the influence of the plasma frequency and vice versa for the BH shadow radii. Also, both of them decrease with the rise of the $\xi$ and $\gamma$ parameters. After that, we constrain the spacetime parameters and the homogeneous plasma frequency by using the observational data for M87* and Sgr A*.

Moreover, we investigate another observational signature of the Schwarzschild-like BH, which is the weak gravitational lensing in the presence of the uniform and non-uniform plasma. We explore the deflection angle for both cases numerically due to the complex spacetime of the considered spacetime, and we find that the deflection angle increases with the increase of the uniform plasma frequency and vice versa for the non-uniform case. The rise of the spacetime parameters leads to a reduction in the values of the deflection angle for both cases.

Finally, we study the magnification of the gravitationally lensed image. Using Einstein's ring, we define the magnification factors for the primary and secondary images, and we find the total magnification by combining them through linear combination. Subsequently, we plot the total magnification for the uniform and non-uniform cases as a function of the plasma frequency and the $\xi$ parameter. The results indicate that there is a decrease with the increase of the $\xi$ and $\gamma$ parameters, and the total magnification decreases under the influence of the non-uniform plasma frequency and for the uniform case, vice versa. To elaborate further, we plot the ratio of the magnification factors in the presence of the plasma to those in the vacuum case. Notably, the values of this ratio increase/decrease under the influence of the uniform/non-uniform plasma frequency.

It is crucial to investigate the optical phenomena around BHs, including the Schwarzschild-like BH, from an astrophysical perspective. Our theoretical findings do not advance fundamental understanding, but may open a window for future observational and experimental efforts to distinguish the Schwarzschild-like BHs from the ordinary Schwarzschild BH.

\section*{ACKNOWLEDGEMENT}

This research was funded by the National Natural Science Foundation of China (NSFC) under Grant No. U2541210.

\bibliographystyle{apsrev4-2}  
\bibliography{Ref}

\end{document}